\documentstyle[prl,aps,twocolumn,epsf]{revtex}
\begin{document}
\title
{\large \bf Boundary conditions at the mobility edge}
\author{D. Braun$^1$, G. Montambaux$^2$ and M. Pascaud$^2$}
\address{$^1$Universit\"at-Gesamthochschule Essen, Fachbereich 7 - Physik,
45117 Essen, Germany\\
$^2$Laboratoire de Physique des
Solides,  associ\'e au CNRS \\
Universit\'{e} Paris--Sud, 91405 Orsay, France}
\twocolumn[
\maketitle
\widetext
\centerline{\today}
\begin{abstract}
\begin{center}
\parbox{14cm}{
 It is shown
that the universal behavior of the spacing distribution
of nearest energy levels at the metal--insulator Anderson transition is
indeed dependent on the
boundary conditions. 
The spectral rigidity $\Sigma^2(E)$ also depends on the
boundary conditions but this dependence vanishes at high energy
$E$.
This implies that the multifractal
exponent $D_2$ of 
the participation ratio of wave functions in the bulk
is not affected by the boundary  conditions.}
\end{center}
\end{abstract}]

\narrowtext

The spectral analysis of disordered conductors has been proven
 recently to be a useful tool to probe the nature of the 
eigenstates\cite{Shklovskii93,Kravtsov94a,Aronov94,Kravtsov95}. 
In the diffusive (metallic) regime, the conductance $g$ scales linearly with
the size $L$ of the system, 
and the wave functions are delocalized over the
sample. The spectral correlations have been shown to be those of random
gaussian matrices\cite{Efetov83},
with large deviations above the Thouless energy $E_c=\hbar/\tau_D= \hbar
D/L^2$\cite{Altshuler86}, where 
$D$ is the diffusion coefficient and $\tau_D$ is the time needed for a
wave packet to cross the sample. 
In particular, the distribution $P(s)$ of
spacings between nearest levels
is very well fitted by the Wigner--surmize characteristic
of chaotic systems \cite{Mehta91}:
$P(s) = (\pi / 2) s \ \exp[- (\pi /4) s^2]
$
where $s$ is written in units of mean level spacing
 $\Delta$\cite{Wigner}. These  deviations of  
order $1/g^2$\cite{Kravtsov94b}   become negligible in the limit of large
$L$.
 In the localized phase, in the limit $L \rightarrow \infty$,
 the levels become completely uncorrelated and $P(s)$ has the
poissonian form:
$P(s) =  \exp(-  s)$.
This is because two levels close in energy are  distant in space
so that their wave functions do not overlap.

It has been found that  the Anderson metal--insulator
transition in three dimensions is characterized by a third
distribution\cite{Shklovskii93}
which
has the remarkable property to be universal, i.e.~it is independent of the
size, whereas 
it is size dependent in the localized and metallic regimes. The transition
is thus described 
as an unstable fixed point, in the sense that slightly above the
transition ($W > W_c$, where
$W$ is the disorder strength and $W_c$ is the critical disorder) 
the distribution
tends to a poissonian limit when $L \rightarrow \infty$, while slightly below
the transition  
($W < W_c$), it tends to the Wigner-Dyson (WD) distribution.
This third universal distribution has been extensively studied by several
groups who 
confirmed these results, for $L$ ranging from $6$ to
$100$
\cite{Altshuler88,Hofstetter93,Evangelou94,Zharekeshev95,Varga95,Braun95,Zharekeshev97}.
Up to now, the  form of the distribution is still unexplained.

 $P(s)$ carries information on the short range part of the
spectral 
correlations. Other characterizations are
the two-level  correlation function (TLCF) of the density of states
$\rho(\varepsilon)$: $K(s)=\langle \rho(\epsilon+s)\rho(\epsilon)
\rangle/\langle \rho(\epsilon) \rangle^2 -1$ and
 the so-called
number variance $\Sigma^2(E)=\langle N^2(E) \rangle -\langle N(E) \rangle
^2$
 which measures the fluctuation of the number of levels $N(E)$ in a band of
width $E$.  
 $E$ is in units of $\Delta$.
It is related to the
TLCF: $\Sigma^2(E)=2 \int_0^E (E-s) K(s) d s\,.$
Surprisingly enough, the numerical studies which lead to the same
shape of the distribution $P(s)$
at the transition 
have apparently been all performed  using {\it periodic boundary
conditions}.  In this work, we
have calculated $P(s)$ at the transition for the same hamiltonian, {\it 
with different boundary
conditions} (BCs).
The hamiltonian is taken as:
\begin{equation}
H=\sum_{i} \varepsilon_i \ c^+_i c_i - t \ \sum_{(i,j)} (c^+_i c_j + c^+_j
c_i) \,.
\end{equation}
The sites $i$ belong to a 3D cubic lattice. Only transfer $t$ between
nearest neighbours $(i,j)$
is considered. The site energies $\varepsilon_i$ are chosen independently
from  a symmetric 
box distribution of width $W$.
The metal--insulator transition occurs at the center of the band for the
critical value 
$W_c=16.5 \pm .2$
\cite{Shklovskii93,Kramer94,Hofstetter93,Zharekeshev95}.
We have found that {\it although the level statistics at the transition
is   independent of the size
of the system, it depends on the boundary conditions}.
 Our main result is shown in Fig.1 where
we have plotted the 
spacing distribution for four types of BCs,
a) periodic in the three directions (the situation
studied by previous authors and that we will refer to as $(111)$, b) periodic
in two directions and "Hard Wall" (HW) (Dirichlet) in
the third $(110)$, c) periodic in one direction and HW in the two others
$(100)$, d) HW in the three directions $(000)$.
All these distributions are "universal"
in the sense
that they are size independent. 
 The critical point
depends at most very weakly on the choice
of the BCs. It seems to shift slightly to smaller $W$ when the number of HW
boundaries is increased. Using standard scaling analysis of $\langle
s^2\rangle $ and $\int_0^2 P(s)\,ds$, we  found $W_c=16.0 \pm .5$ for the
(000) geometry. However, within the range of sizes studied ($L=12,\ldots,
22$)
the difference between the $P(s)$ at $W=16.0$ and at $W=16.5$ is negligible
compared to the remaining statistical fluctuations of the spacing
distribution.

In Fig.2 we have plotted the second moment of
the level spacing
$\langle s^2 \rangle$ as a function of the size for the different BCs.
This plot shows that the distributions are size independent and that they do
not converge to a single one in the large size limit.

It may appear  {\it a priori} surprising that the distribution is  at the
same time 
size independent and sensitive to the BCs.
To clarify this point, it is instructive to recall
the behavior of
the typical dimensionless curvature $g_d=\pi \langle |c| \rangle / \Delta$
of the energy levels when an infinitesimal
flux is introduced in the cylinder geometry.
In the metallic regime,  
$g_d(L)$ increases linearly with
the size and it decreases   exponentially
in the localized regime. At the  transition, the curvature $g_d(L)=g^*_d$ is
size 
independent\cite{Canali96,Braun97}. Since $g_d$ measures the sensitivity of
energy
levels to a change of the BCs, the simple fact that it is non-zero shows
that the spectral correlations can be at the same time size independent
and sensitive to the  BCs.
 This universal sensitivity to the BC has already been discussed in the
case of {\it periodic BCs} where one or several Aharonov-Bohm (AB) fluxes
were applied\cite{Batsch96,Montambaux97}.  However in that case, the
 symmetry --- time reversal invariance ---
was at the same time broken by the fluxes,
such that it is not
surprising that the statistics is changed.

The distribution found by other authors with periodic BCs is
the most rigid of the four distributions that we have studied.
When periodic BCs are relaxed and replaced by hard wall BCs, the distribution
becomes closer to the Poisson distribution, with a short range repulsion
which is characterized by a larger slope of $P(s)$. 
The slope $P'(0)$ varies by more than a factor three from
$2.14$ \cite{Shklovskii93} for the BC $(111)$ to $6.80$ 
for the BC $(000)$ (see table I).

It is useful to stress that in the metallic regime itself, there are
deviations to the WD distribution which depend on the BCs.
These deviations are related to a contribution of the diffusive
modes\cite{Kravtsov94b}. At small $s$, the slope of $P(s)$ is given by
\begin{equation}
P(s)={\pi^2 \over 6} (1 + {3 a \over  \pi^6 g^2})  \ s\,,
\end{equation}
where the coefficient $a$
describes the diffusive motion and is
given by
\begin{equation}
a={\pi^4 \over L^4} \sum_{q \neq 0} {1 \over ({\bf q}^2)^2 }\,.
\end{equation}
For an isolated system, the diffusion modes are quantized by the BCs.
In a direction where the boundaries are hard walls, $q=n \pi / L$ with
$n=0,1,2,3,\cdots$.
 In a direction where the boundaries are periodic, $q=2 n \pi /
L$ with $n=0,\pm 1,\pm 2,\pm 3,\cdots$.  In $d=3$, one finds:
$a_{111}=1.03 \ , \ a_{110}=2.15  \ , \ a_{100}=3.39
  \ , \ a_{000}=5.13\,.$
So in a metal, the slope of $P(s)$ depends on the BC. However,
the corrections are order $1/g^2$
and decrease with the size  since $g(L) \sim L$, and they vanish for
the infinite system.
At the
transition $g=g^*$ is size independent and one may expect that the
correction to $P(s)$ still depends on the BC through the quantization of
the anomalous diffusion modes.
This correction can be also simply calculated for an
anisotropic system.
It depends on the shape of the sample. This certainly means that the spectral
correlations at the transition are also shape dependent\cite{Potempa98}.

The distributions
 we have found  bear an interesting similarity with another recently
 studied distribution\cite{Bogomolny97}. 
A remarquable and simple spectral sequence which is intermediate between the
WD and the Poisson 
distributions is obtained by taking the middles of a poissonian
sequence. This new sequence  
has been baptized ``semi-Poisson''\cite{Bogomolny97}. The corresponding
$P(s)$ is   given by\cite{Bogomolny97,Pascaud97}:
\begin{equation}
P(s)= 4 s e^{- 2 s}
\label{semipoisson}\,.
\end{equation} 

It has been shown that the 
equilibrium
distribution of charges
in a Coulomb gas with logarithmic interaction only between {\it nearest
neighbors} 
is also described by eq.(\ref{semipoisson}). The TLCF
and $\Sigma^2(E)$  for this model (referred to later
as Short Range Plasma Model, SRPM) are however different from those for
 the semi--Poisson sequence. 
 We shall return to
this point later.

In the inset of Fig.3 we have plotted the arithmetic average of the four
distributions. Quite amazingly, it is very close to
the semi--Poisson distribution. The average of the slope
at small separation calculated with the four BCs is 
$4.08\pm 0.4$
instead of $4$ for 
the semi--Poisson.
As another characteristic of $P(s)$, the second moment
$\langle s^2 \rangle$ is shown in table I for the various BCs.
The average over the different BCs is found to be 
$1.51\pm 0.01$.
It is $3/2$  for the semi--Poisson.
The tails of  $P(s)$ have also been considerably
studied
\cite{Shklovskii93,Aronov94,Kravtsov95,Evangelou94,Varga95,Zharekeshev97}. 
The inset of Fig.1 shows the tails for the four BCs. They clearly differ by the
rapidity of their decay, the usual periodic BCs giving rise to the fastest
decay.
It is interesting to notice that the behavior at large $s$ is
much more
affected by 
HW BCs than by a simple addition of a AB flux or even a
magnetic field\cite{Batsch96}.

 We have also investigated random
boundary conditions,  with  random
hopping terms $t_{ij}=t_{ji}$
connecting opposite sides of the sample. Drawing the $t_{ij}$ for
each disorder realization 
independently from a box distribution centered around zero and with
width $\tau$,
 we found a
continuous family of ``universal'' critical ensembles, which are for finite
$\tau$ distinct from the ones with ``deterministic'' boundary conditions.

We now turn to the number variance. A linear behavior at  large
$E$,
$\Sigma^2(E)/E \rightarrow \chi$ defines  the level
compressibility $\chi$, which is also related to the $t \rightarrow 0$
dependence of the form factor $\tilde K(t)$, the Fourier transform of
$K(s)$.
One has $\chi = \int_{-\infty}^\infty K(s) ds
=\tilde K(0)\,.$
This means $\chi=1$ for the Poisson and semi-Poisson sequences, $\chi=0$ for
 the WD correlations and 
$\chi=1/2$ for the SRPM.

In Fig. 4, we have plotted
$\Sigma^2(E)/E$ for the various BCs. It is
seen that, like for $P(s)$, the rigidity depends on the BCs for small energy
ranges. The rigidity is weaker for
non-periodic
BCs. 
However, when $E$ increases, the different rigidities seem to converge
towards the same value
(see inset of fig.4). We find $\chi \simeq 0.27 \pm
0.02$ in agreement  with  previous
 authors\cite{Altshuler88,Zharekeshev94}.  Within error bars, this
asymptotic value does not depend on the BCs.

$P(s)$ and $\chi$ carry information on different time
scales in the problem.
Let us remind that the metallic spectrum is characterized by two
characteristic time scales, the Thouless time $\tau_D$ and the Heisenberg
time $\tau_H=h/\Delta$,
with $\tau_H / \tau_D=2 \pi E_c/\Delta \gg 1$. At the transition, these two
times are of the same order. Consequently,  correlation functions like
$P(s)$
which probe correlations at energy scales of the order of $\Delta$, i.e. time
scales  of the order of 
$\tau_H \simeq \tau_D$, probe
the sensitivity to the boundary conditions of a wave packet evolution.
However, the asymptotic form of the spectral rigidity at energy $E \gg
\Delta$ typically probes time scales $t \ll t_H \simeq \tau_D$ for which the
diffusion of a wave packet is insensitive to the  BCs.

More precisely, the form factor  has been shown to be
related to the return  probability $P(t)$ for a wave
packet\cite{Argaman93,Chalker96}.  At the transition, the wave
functions have a multifractal
structure\cite{Castellani86,Schreiber85}, with a long range tail
showing a power law decay.
Multifractality is characterized by the time dependence, 
when $t \ll
\tau_D$: \begin{equation}P(t) \propto t^{-D_2/d}\,,
\label{Pt}
\end{equation}
 where $D_2$ is the multifractal 
exponent of the inverse participation
ratio, $\sum_{\bf r} \langle | \psi_n({\bf r})|^4 \rangle
\propto L^{-D_2}$\cite{Castellani86}.
 From the limit $t \rightarrow 0$ of $P(t)$, $\chi$ is found to
be\cite{Chalker96}: 
$$\chi={1 \over 2}(1 - {D_2 \over d})\,.$$

The multifractal exponent
 $D_2$ defined from small time behavior is thus 
expected to be independent of the BCs. Therefore  $\chi$ should not
depend on the BCs either.
This is seen to be true from
the inset of fig. 4, where we show that $\Sigma^2(E)/E$ converges to the
same value $\chi \simeq 0.27 \pm 0.02$ for periodic and HW BCs.  
This value of $\chi$ leads to  a multifractal
 exponent $D_2 \simeq 1.4\pm 0.2$ which has to be compared
with the value $D_2 \simeq 1.6 \pm .3$ found from other direct numerical
calculations  involving the study of the wave
functions \cite{D2}.

In conclusion, we have shown that the spectral correlations at the
metal--insulator transition, although being universal, i.e. independent of
the size, strongly depend on the choice of the boundary
conditions. This dependence is most pronounced
 for small energy scales.
When the periodic BCs are replaced by hard walls in one or more directions,
the spectrum becomes less 
and less rigid.
However, the level compressibility defined from the $E \rightarrow \infty$
limit of the spectral rigidity is
independent of the choice of the boundaries. 

We acknowledge stimulating discussions with E. Bogomolny, V. Falko, I.
Lerner, F. Pi\'echon, and P. Walker. Numerical simulations have been
performed using IDRIS facilities (Orsay).
G. M. thanks the Isaac Newton Institute for Mathematical Sciences
for hospitality during the completion of this work.

 \bigskip

\begin{center}
\begin{tabular}{|c||c|c|c|c|c|c|c|}   \hline
		      & Wigner    & 111  & 110 & 100 & 000 & SRPM
& Poisson\\  \hline
$P'(0)$                     &${\pi^2 \over 6}= 1.65$  & $2.14^{(2)}$ &
$3.01$     
&$4.37$ &$6.80$ & 4        & $\infty$ \\
$\langle s^2 \rangle$ &${4 \over \pi}^{(1)}= 1.27$ & $1.41^{(3)}$ &
$1.48$
&$1.55$ &$1.62$ & 3/2        & 2    

\\ \hline
\end{tabular}\end{center}

Table I :   Numerical results for the various 
measures of spectral correlations compared with the 
SRPM. The relative errors (standard deviations from 6 system sizes,
$L=12,14,16,18,20,22$;
500 to 33 disorder realisations) for $B$ are always less than 10\% and for
$s^2$ less than 1\%. 
$^{(1)}$ This is the value deduced from the Wigner surmize;
$^{(2)}$ See also ref.\cite{Shklovskii93};
$^{(3)}$ See also ref.\cite{Varga95}.

\begin{figure}
\centerline{ \epsfxsize 8cm
\epsffile{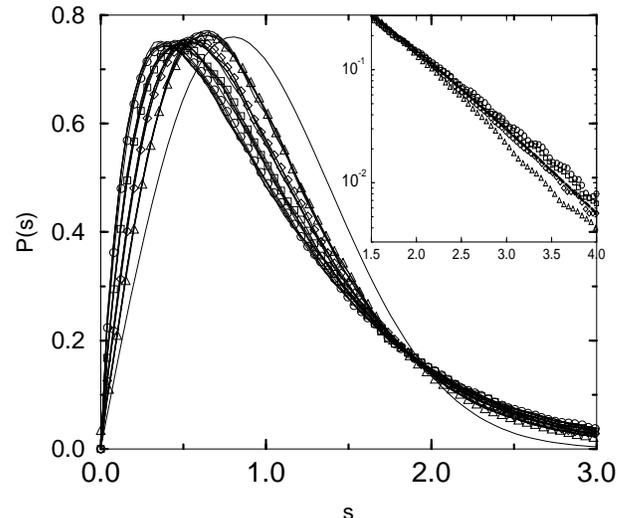}}
\caption{Distribution $P(s)$  at the metal-insulator
transition 
with four different types of boundary conditions defined in the text:
$\protect{\small \bigtriangleup}$ 
$111$, $\protect\diamond$ $110$, $\protect{\small \Box}$ $100$, and
$\protect\circ$ 
$000$. Distributions with $L=8$ to $L=14$ are shown.
The  Wigner--Dyson result (continous line) is also plotted. 
In the inset the tails of $P(s)$ are shown for $L=10$ and compared with
the Semi-Poisson distribution, eq.(4) (dashed line).} \end{figure}

\begin{figure}
\centerline{ \epsfxsize 8cm \epsffile{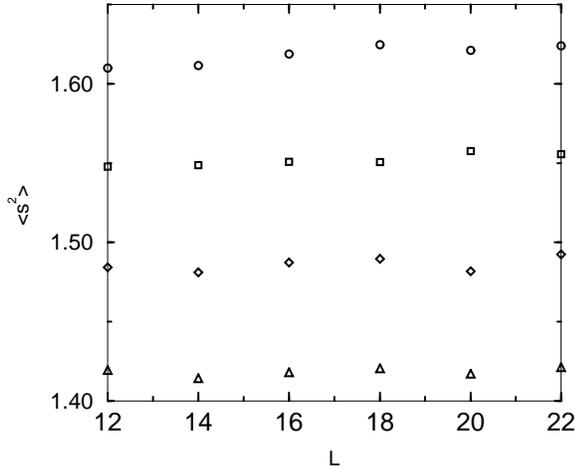}}
\caption{$\langle s^2 \rangle$ versus linear size $L$ for different BCs
(symbols as in Fig.1) 
 for $W=16.5$.       }
 \end{figure}

 \begin{figure}
\centerline{ \epsfxsize 7cm \epsffile{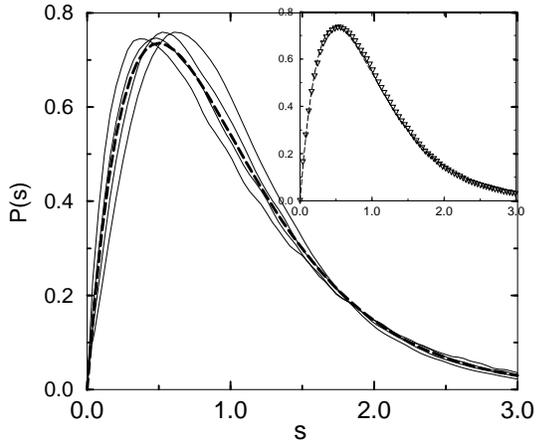}}
\caption{$P(s)$ for the four different BCs
compared with semi--Poisson (dashed line).
Inset:``Average''
$\protect P(s)$ at the transition
($\protect\small\bigtriangledown$), compared
with semi--Poisson.} \end{figure}

\begin{figure}
\centerline{ \epsfxsize 8cm \epsffile{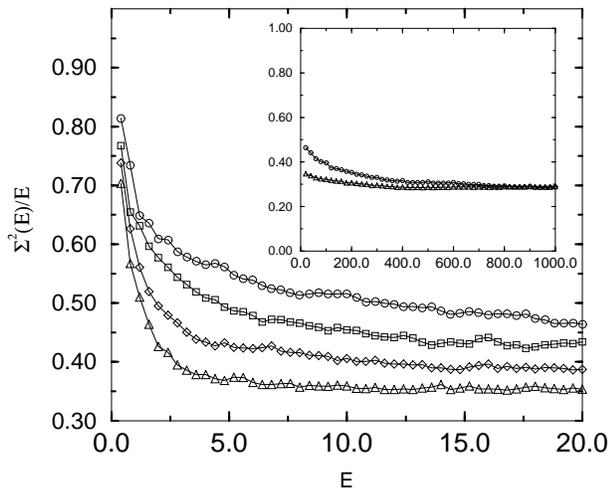}}
 \caption{$\Sigma^2(E)/E$ for the different
 BCs (average over $L=20$ and $22$). 
Symbols as in Fig.1.
The inset
shows that at large energy the difference between the BCs (111) and (000)
vanishes. }
\end{figure} 


\begin{thebibliography}{99}
\bibitem{Wigner}This simple surmize derived for a $N$  x $N$ random
matrix with $N=2$ is a rather good approximation
of the $N = \infty$ result. The slope at the origin is $\pi^2/6$ for $N =
\infty$.
\bibitem{Potempa98} During the revision of this paper, we have been aware
that a recent work
has reached this conclusion, H. Potempa and L. Schweitzer, cond-mat/9804312
\end{thebibliography}
\end{document}